\documentclass[aps, pre, floatfix]{revtex4}
\usepackage{amsmath}
\usepackage{verbatim}
\usepackage[english]{babel}
\usepackage[T1]{fontenc}
\usepackage{graphicx}
\usepackage{listings}
\usepackage{lscape}
\usepackage{makeidx}
\usepackage{url}
\usepackage{subfigure}
\usepackage{array}
\usepackage{amsfonts}
\usepackage{verbatim}

\newcommand{\lpar}{\left(}
\newcommand{\rpar}{\right)}
\newcommand{\lbk}{\left \lbrack}
\newcommand{\rbk}{\right \rbrack}

\newcommand{\ga}{{\alpha}}
\newcommand{\gb}{{\beta}}
\newcommand{\gc}{{\gamma}}
\newcommand{\gd}{{\delta}}
\newcommand{\gab}{{\alpha \beta}}
\newcommand{\gac}{{\alpha \gamma}}
\newcommand{\gbc}{{\beta \gamma}}
\newcommand{\gca}{{\gamma \alpha}}

\newcommand{\pt}{\partial_t}
\newcommand{\pa}{\partial_\alpha}
\newcommand{\pb}{\partial_\beta}
\newcommand{\pc}{\partial_\gamma}
\newcommand{\beq}[1]{\begin{equation}\label{#1}}
\newcommand{\eeq}{\end{equation}}
\newcommand{\eref}[1]{Eq.~(\ref{#1})}

\begin{document}
%\title{Fundamental Theory of Multi-Relaxation Time Lattice Boltzmann Methods}
%\title{General Asymptotic Derivation of Hydrodynamics for Lattice Boltzmann}
\title{Derivation of Hydrodynamics for Multi-Relaxation Time Lattice Boltzmann using the Moment Approach}

\author{G.~\surname{Kaehler}}
\email{goetz.kaehler@ndsu.edu}
\author{A.~J.~\surname{Wagner}}
\email{alexander.wagner@ndsu.edu}
\affiliation{Department of Physics, North Dakota State University, Fargo, ND 58108, U.S.A.}
%\author[G.~Kaehler]{Goetz~Kaehler\affil{1}\corrauth and Alexander~Wagner\affil{1}}
%\address{\affilnum{1}

%\emails{{goetz.kaehler@ndsu.edu} (G.~Kaehler)}

\begin{abstract}
\noindent A general analysis of the hydrodynamic limit of multi-relaxation time lattice Boltzmann models is presented. We examine multi-relaxation time BGK collision operators that are constructed similarly to those for the MRT case, however, without explicitly moving into a moment space representation. The corresponding 'moments' are derived as left eigenvectors of said collision operator in velocity space. Consequently we can, in a representation independent of the chosen base velocity set, generate the conservation equations. We find a significant degree of freedom in the choice of the collision matrix and the associated basis which leaves the collision operator invariant. Therefore we can explain why MRT implementations in the literature reproduce identical hydrodynamics despite being based on different orthogonalization relations.
%In fact, we show that these methods are identical wi 
% one can construct the commonly used collision matrices of different MRT implementations, which have eigenvectors orthogonal with respect to different inner product. Despite their formal difference, these different norms and thus different types of the collision matrices are shown to be identical with respect to their effect on the collision operator.
\end{abstract}
\maketitle

\section{Introduction}
The lattice Boltzmann (LB) method is continuing to increase in popularity as a simulation method for fluid mechanics for a wide range of applications from turbulence \cite{eggels-1996} to complex fluids \cite{shan-1993}. A key of its success is the simplicity of the algorithm. Instead of discretizing the hydrodynamic equations directly the method is based on an underlying microscopic model. Historically the method developed from lattice gases \cite{frisch-1986} where particles move on a lattice and collide on lattice points. Because such a lattice gas model locally conserves mass and momentum the macroscopic behavior of the system has to be described by the continuitiy and Navier-Stokes equations \cite{succi-2001}. The connections between the microscopic streaming and collision rules and the macroscopic differential equations is established by taking the hydrodynamic limit which requires averaging the locally conserved quantities. This reproduces the Boltzmann equation \cite{kerson-1987}. Performing a Taylor expansion on the discrete Boltzmann equation then leads to a PDE representation of the discrete evolution equation \cite{dellar-2002}.

At this point there are several routes to proceed. Grad \cite{grad-1949} suggests taking moments of the full Boltzmann equation which is a route that has been taken by other groups \cite{swift-1996}. Alternatively one can formally expand the distribution function before taking the moments, which is known as the Chapman-Enskog expansion \cite{giraud-1997}. Either approach will lead to identical results to second order: the continuity and Navier-Stokes equations as well as the heat equation for thermal systems. The higher order equations are, however, quite different. Here neither approach has been particularly successful as the Navier-Stokes level equations appear to be appropriate to length-scales close to molecular scale \cite{koplik-1995}. There are few attempts to derive higher order hydrodynamic equations in the LB context. One exception are multi-phase fluids where higher order spatial derivatives giving rise to surface tension have to be taken into account \cite{wagner-2006-2}.

The development of the method took a major leap when it was discovered that it is feasible to use a Boltzmann--level microscopic model \cite{mcnamara-1988}, which removes microscopic noise. This approach is referred to as the lattice Boltzmann method. Qian {\em et al.} \cite{qian-1992} found that the approach is simplified considerably when the collision operator is written as a single-time BGK expression which relaxes local particle distributions towards the equilibrium distribution. To this date this represents the most popular flavor of lattice Boltzmann algorithms employed.

Shortly after the introduction of the single-time relaxation collision operator d'Humieres realized that one can extend the BGK collision with a multi-relaxation time (MRT) approach \cite{dhumieres-1992}. This replaces the scalar relaxation time with a matrix, which then allows for decoupled relaxation of different stress terms. Thus it decouples the different transport coefficients and they no longer need to take their ideal gas values. Up to now deriving hydrodynamic equations for multi-relaxation time lattice Boltzmann methods has been almost exclusively achieved by Chapman-Enskog like expansions. These expansions typically depend on the specific model \cite{giraud-1997} although a very general (but not very detailed) approach was presented by Junk {\em et al.} \cite{junk-2005}. Here we show that it is straight forward to derive the hydrodynamic equations through a moment approach in a model-independent manner. Our approach clarifies the structure of the collision operator by showing that the stress moments have to be left eigenvectors of the collision matrix. This feature has been obscured by current implementations of multi-relaxation time LB algorithms, which, either to reduce the degrees of freedom in the choice of the collision matrix \cite{dhumieres-1992} or because of additional properties required for analysis of the method \cite{benzi-1992, adhikari-2005}, choose that the eigenvectors of the collision matrix be orthogonal with respect to a specific scalar product.

It may be initially surprising that the freedom to choose orthogonality with respect to different scalar products exists. The underlying reason, however, is simple: because the conserved moments are left invariant by the collision, these moments can be mixed into the non-conserved moments without changing the collision operator. With this observation we then reconcile the existing multi-relaxation time lattice Boltzmann models with the analysis in this paper.

\section{Lattice Boltzmann}
The lattice Boltzmann equation~(LBE) is a representation of the Boltzmann transport equation~\cite{kerson-1987} with three levels discretization taken into account: time~$t$ , position~${\bf x}$ and velocity~${\bf v}$. First LB-methods utilized a two body collision operator derived from lattice gas methods (Higuera {\em et al.} \cite{higuera-1989}). Later Qian and D'Humieres realized that the collision operator could be significantly simplified using a BGK approach \cite{bhatnagar-1954} as $\Omega_i=(1/\tau) (f_i^0-f_i)$ where $f_i^0$ is the local equilibrium distribution ~\cite{dhumieres-1992, qian-1992}. 
 In the BGK approximation~\cite{bhatnagar-1954} the collision integral is replaced by a relaxation term that moves the current distribution $f(\mathbf{x}, \mathbf{v}, t)$ function towards the equilibrium distribution $f^0(\mathbf{x}, \mathbf{v}, t)$
%$f^0(\rho, \mathbf{u}, \theta)$. 
For a general collision operator $\Omega_i$ the basic LBE can then be written as
\begin{equation}
\label{eqn:lbe}
f_i(\mathbf{x}+ \mathbf{v}_i, t+1) - f_i(\mathbf{x}, t) = \Omega_i(f_1,\cdots,f_N),
\end{equation}
where the $f_i$ are the density functions associated with a discrete set of $N$ base velocity vectors $\mathbf{v}_i$, $x$ is the lattice position and $t$ is the discrete time with interval $\Delta t = 1$. The velocities are chosen such that the $v_i$ are lattice vectors. Since collisions conserve certain quantities such as mass and momentum we require
\begin{equation}
\label{eqn:conservation}
\sum_i \psi_i^{a,c}\Omega_i = 0,
\end{equation}
where the $\psi_i^{a,c}$ are the vectors describing the velocity moments of the conserved quantities. The index $c$ only emphasizes that these vectors are associated with conserved quantities. We will encounter non-conserved vectors $\psi^a$ later in this paper. The first quantity that has to be conserved in the collision is the local density which has a corresponding vector of $\psi^{0,c}_i = 1_i$ where $1_i$ is simply $1$ for every $i$. Momentum must also be conserved in each spatial direction. In three dimensions the corresponding $\psi$ vectors are  $\psi_i^{1,c} = v_{i,x}$, $\psi_i^{2,c} = v_{i,y}$, and $\psi_i^{3,c} = v_{i,z}$.  We denote the locally conserved quantities as density $\rho$ and momentum $\mathbf{j}$. They are defined through the vectors $\psi^{a,c}$ as 
%quantities density $\rho$ and momentum $\mathbf{j}$ can thus be defined as
\beq{eqn:consquant}
%\sum_i \psi_i^{0,c} f_i = \rho, \;\;\; \sum_i \psi_i^{1,c} f_i = j_x, \;\;\; \sum_i \psi_i^{2,c} f_i = j_y, \;\;\;, \sum_i \psi_i^{3,c} f_i = j_z.
\sum_i \psi_i^{0,c} f_i = \rho, \;\;\; \sum_i \psi_i^{\ga,c} f_i = j_\ga.
\eeq
Throughout this paper greek indices $\ga, \gb, \gc$ will generally denote the range of spatial dimensions $\lbrace x, y, z \rbrace$ and be treated under the Einstein summation convention. Latin indices $i, j, k$ are used in the context of vector components of the lattice Boltzmann base velocity set and are summed over explicitly. 

Most LB models are used to simulate isothermal hydrodynamics and these models are the focus of this paper. Thermal models require the conservation of the additional moment $v_i^2$, which we do not treat here. In principle, however, it should be easy to extend the presented approach to thermal systems and generate the corresponding heat equation.
% but as most lattice Boltzmann applications limit themselves to the isothermal case. 
 %For simplicity we will limit ourselves here to the isothermal case.

%First lattice Boltzmann methods utilized a two body collision operator, derived from lattice gas methods (Higuera {\em et al.} \cite{higuera-1989}). Qian and D'Humieres realized that the collision operator could be significantly simplified using a BGK approach as $\Omega_i=(1/\tau) (f_i^0-f_i)$ where $f_i^0$ is the local equilibrium distribution ~\cite{dhumieres-1992, qian-1992}. 

To recover the continuity and Navier-Stokes equations this local equilibrum distribution needs to match the first four velocity moments of the continuum Maxwell-Boltzmann distribution. This distribution is 
%This equilibrium distribution needs to match the
%first few velocity moments of the continuous Maxwell Boltzmann distribution 
$f^0(\mathbf{v})=\frac{\rho}{(2\pi \theta)^{3/2}}
\exp\left(\frac{(\mathbf{v}-\mathbf{u})^2}{2\theta}\right)$ where local velocity is defined as $\mathbf{u} = \mathbf{j}/\rho$ and $\theta$ is the temperature. For thermal models we would need to match velocity moments. 
%We require these moments in the derivation of the hydrodynamic limit. To derive isothermal hydrodynamics we need the first four moments given by
The first four moments sufficient to derive isothermal hydrodynamics are
\begin{eqnarray}
\label{eqn:rho}
\sum_i f_i^0 & = & \rho, \\
\label{eqn:j}
\sum_i v_{i\ga} f_i^0 & = & \rho u_\ga = j_\ga,\\
\label{eqn:v2}
\sum_i v_{i\ga} v_{i\gb} f_i^0 & = & \rho \theta + \rho u_\ga u_\gb, \\
\label{eqn:v3}
\sum_i v_{i\ga} v_{i\gb} v_{i\gc} f_i^0 & = & \rho \theta \lpar u_\ga \delta_\gbc + u_\gb \delta_\gca + u_\gc \delta_\gab \rpar + \rho u_\ga u_\gb u_\gc + Q_{\ga\gb\gc}.
\end{eqnarray}
The tensor quantity $Q_{\ga\gb\gc}$ is an arbitrary correction term and vanishes in the continuum case. However, the typical choice is $Q_{\ga\gb\gc} = - \rho u_\ga u_\gb u_\gc$ which allows us to use a much smaller velocity set. The tradeoff are small Galilean invariance problems~\cite{wagner-2006}. Note that the conserved moments of the local equilibrium distribution $f_i^0$ and the distribution $f_i$ are identical because the collision does not change them, \textit{i.e.} $\sum_i \psi_i^{a,c} f_i^0 = \sum_i \psi_i^{a,c} f_i$.

Depending on the base velocity set the conditions Eqs.~(\ref{eqn:rho}-\ref{eqn:v3}) may not uniquely define the equilibrium distribution. From several different general arguments it is usually found that the explicit form
\begin{equation}
\label{eqn:fnull}
f_i^0(\rho, \mathbf{u}) = \rho w_i \left\lbrack( \mathbf{1} + \frac{1}{\theta} \mathbf{u}.\mathbf{v}_i + \frac{1}{2\theta^2}\left(\mathbf{u}.\mathbf{v}_i\right)^2 - \frac{1}{2\theta}\mathbf{u}.\mathbf{u}\right\rbrack 
\end{equation}
is the optimal choice for the isothermal equilibrium distribution for an appropriate choice of the $w_i$ weight constants \cite{he-1997-a} although other forms have been used \cite{wagner-1999}.

When deriving the hydrodynamic equations from the continuous Boltzmann equation using the single-relaxation time
approximation leads to a fixed ratio of the transport coefficients such as
shear viscosity, bulk viscosity, and thermal conductivity \cite{kerson-1987}. In the discrete case of lattice Boltzmann the same hydrodynamic equations can be derived with transport coefficients containing a renormalized relaxation time $\omega = \lpar \tau - 1/2\rpar$.
For ideal gases the predicted ratios agree quite well with the experimentally
measured values \cite{kerson-1987}. The form of the hydrodynamic equations
apply not only to ideal but also non-ideal gases and even fluids. Lattice Boltzmann applications usually consider examples from this
more general class of systems. In these more general cases, however, the
ratios of transport coefficient are no longer fixed, and it would be
advantageous to write a more flexible collision term that allows for
independently variable transport coefficients. This was accomplished
by D'Humieres \cite{dhumieres-1992} by considering a multi-relaxation time BGK collision
operator of the form
\begin{equation}
\label{eqn:mrtlbe}
\Omega_i(f_1,\cdots,f_N) = \sum_j \Lambda_{ij} \lbrack f_j^0(\rho, \mathbf{u}, \theta) - f_j(\mathbf{x}, t)\rbrack, 
\end{equation}
where $\Lambda$ is a collision matrix. If we choose
$\Lambda_{ij}=\delta_{ij}/\tau$ we recover the single-relaxation time
collision operator. Another numerical rationale for implementing multi-relaxation time Lattice Boltzmann methods is the improvement in stability, particularly for high Reynolds numbers \cite{lallemand-2003}. 
There are some requirements on the collision matrix to ensure mass and momentum conservation in the collision. 
In the single-relaxation time approach the conservation laws were respected because the conserved moments of the local distribution $f_i$ and the
local equilibrium distribution $f_i^0$ are identical. For the
multi-relaxation time collision term~\eref{eqn:conservation} requires
\beq{eqn:cons2}
\sum_i \psi_i^{a,c} \sum_j \Lambda_{ij} \left(f_j^0-f_j\right) = 0.
\eeq
These equations will be satisfied if we demand that the scalar product of a conserved quantity vector with the collision matrix is a linear combination of conserved quantity vectors, \textit{i.e.} $\sum_i \psi_i^{a,c} \Lambda_{ij} = c^a \psi_j^{a,c}$ for an arbitrary $c^a$. Note here that the only physically relevant quantity is the collision operator $\Omega$, not the collision matrix $\Lambda$. While different choices for $c^a$ will lead to different collision matrices, they will not change the collision operator $\Omega$. A convenient choice that coincides with the single-relaxation time case sets the conserved moments $1_i$ and $v_{i\alpha}$ to the left-eigenvectors of our collision matrix with some eigenvalue:
\begin{eqnarray}
\label{eqn:rhoev}
\sum_i 1_i \Lambda_{ik} &=&  \frac{1}{\tau_\rho}1_k,\\
\label{eqn:jev}
\sum_j v_{j\ga} \Lambda_{ji} &=& \frac{1}{\tau_{i_\ga}} v_{i\ga},
\end{eqnarray}
where we used the relaxation times $\tau$ to denote the inverse
eigenvalues of the collision matrix. This choice also allows us to ensure that $\Lambda$ is invertible which, while not strictly necessary, simplifies the formalism. Clearly, the values of $1/\tau_\rho$ and $1/\tau_{j_\ga}$ are entirely arbitrary, meaning that $\tau_\rho$ and $\tau_{j_\ga}$ may not appear in the hydrodynamic equations.

\section{Hydrodynamic limit by the moment method}
In this section we present a new approach to obtain the hydrodynamic equations for the multi-relaxation time lattice BGK equation. We generalize the moment approach familiar from single-relaxation time methods~\cite{wagner-2006} to the more general MRT formalism. For the multi-relaxation time collision operator we expand the left hand side of~\eref{eqn:lbe} to second order:
\beq{eqn:lbeexp}
\lpar \pt + v_{i\ga} \pa \rpar f_i + \frac{1}{2} \lpar \pt + v_{i\ga} \pa \rpar \lpar \pt + v_{i\gb} \pb \rpar f_i + O(\partial^3) = \sum_j \Lambda_{ij}\lpar f_j^0 - f_j \rpar.
\eeq
This allows us to write the $f_i$ in terms of the $f_i^0$ and higher order derivatives as long as $
\Lambda^{-1}$ exists:
\beq{eqn:lbe1st}
f_j = f_j^0 - \sum_i \lpar\Lambda^{-1}\rpar_{ji} \lbk \lpar \pt + v_{i\ga} \pa \rpar  f_i \rbk + O(\partial^2).
\eeq
This is important because we can express the equilibrium distributions $f_i^0$ in terms of $\rho$ and $\mathbf{u}$ in~\eref{eqn:fnull} but not the local distributions $f_i$. Here we have made the assumption that both, spatial and temporal derivatives, are small quantities of the same order of magnitude.
%It should be noted that throughout this paper we assume the differentials for spatial $\pa$ and temporal $\pt$ derivatives are small compared to the discretization intervals.
% Furthermore we abbreviate derivatives as $\frac{\partial}{\pt} = \pt$ and $\frac{\partial}{\pa} = \pa$,~\textit{i.e.} Here we abbreviate the time derivative and
As a byproduct we see that the conservation equations by virtue of~\eref{eqn:cons2} and the $\psi_i^{a,c}$ being left-eigenvectors of $\Lambda_{ij}$ require
\beq{eqn:lbe22}
\sum_j \psi_j^{a, c} \sum_i \lpar \Lambda^{-1} \rpar_{ji} \lbk \lpar \pt + v_{i\ga} \pa \rpar f_i\rbk = \sum_i \tau^a \psi_i^{a,c} \lbk \lpar \pt + v_{i\ga} \pa \rpar f_i\rbk = O(\partial^2),
\eeq
which we will use later. Replacing all occurances of $f_i$ in \eref{eqn:lbeexp} with \eref{eqn:lbe1st} up to second order we obtain
\beq{eqn:lbeexp2}
\lpar \pt + v_{j\ga} \pa \rpar f_j^0 - \lpar \pt + v_{j\ga} \pa \rpar \sum_i \lbk \lpar \Lambda^{-1}\rpar_{ji} - \frac{1}{2} \delta_{ji} \rbk \lpar \pt + v_{i\gb} \pb \rpar f_i^0 + O(\partial^3) = \sum_i \Lambda_{ji} \lpar f_i^0 - f_i \rpar.
\eeq
Because we know $f_i^0$ as a function of $\rho$ and $\mathbf{j}$ this is an equation expressed entirely in terms of our hydrodynamic variables, except for the collision term. 
So far the only requirement on the collision Matrix $\Lambda$ is that it be invertible and fulfill~\eref{eqn:cons2}.
The general approach now to obtain a conservation equations is to take the inner product of the conserved quantity vectors $\psi^{a, c}$ with~\eref{eqn:lbeexp2}. The collision term then vanishes, we retain no dependencies on the $f_i$, and, after some algebra, we obtain the conservation equations. 
%It is useful to take a closer look at the second order derivative term in~\eref{eqn:lbeexp2}
%\beq{eqn:lbe2ot}
%\pt \sum_i \lbk \lpar\Lambda^{-1}\rpar_{ji} - \frac{1}{2} \gd_{ji} \rbk
%\eeq

\subsection{The Continuity Equation}
To obtain the continuity equation we take the inner product of $\psi^{0, c}_j = 1_j $ with~\eref{eqn:lbeexp2} from the left hand side, \textit{i.e.} we just sum over \eref{eqn:lbeexp2} while making use of mass conservation in~\eref{eqn:cons2}. We get
%\beq{eqn:cont1}
% \sum_k \lpar \pt + v_{k\ga} \pa \rpar f_k^0 - \sum_k \sum_i \lbk \lpar\Lambda^{-1}\rpar_{ki} - \frac{1}{2} \delta_{ki} \rbk \lpar \pt + v_{i\ga} \pa \rpar \lpar \pt + v_{k\gb} \pb \rpar f_i^0 + O(\partial^3) = 0.
%\eeq
%\sum_k \sum_i \Lambda_{ki} \lpar f_i^0 - f_i \rpar = 0
\beq{eqn:cont1}
\sum_j 1_j \lpar \pt + v_{j_\ga} \pa \rpar f_j^0 - \sum_j 1_j \lpar \pt + v_{j\ga} \pa \rpar \sum_i \lbk \lpar \Lambda^{-1}\rpar_{ji} - \frac{1}{2} \delta_{ji} \rbk \lpar \pt + v_{i\gb} \pb \rpar f_i^0 + O(\partial^3) = 0 .
\eeq
We can rewrite the second order terms as
\beq{eqn:cont1a}
\pt \sum_j 1_j \sum_i \lbk \lpar \Lambda^{-1}\rpar_{ji} - \frac{1}{2} \delta_{ji} \rbk \lpar \pt + v_{i\gb} \pb \rpar f_i^0 = O(\partial^3),
\eeq
\beq{eqn:cont1b}
\pa \sum_j 1_j v_{j\ga} \sum_i \lbk \lpar \Lambda^{-1}\rpar_{ji} - \frac{1}{2} \delta_{ji} \rbk \lpar \pt + v_{i\gb} \pb \rpar f_i^0 = O(\partial^3),
\eeq
where we used that both $1_j$ and $1_j v_{j\ga} = v_{j\ga}$ are conserved quantity vectors so that we can apply~\eref{eqn:lbe22}.
We are left with
\beq{eqn:cont4}
\sum_j \lpar \pt + v_{j_\ga} \pa \rpar f_j^0 + O(\partial^3) = 0,
\eeq
which using \eref{eqn:rho} and \eref{eqn:j} becomes the continuity equation
\beq{eqn:contfinal}
\pt \rho + \pa \lpar \rho u_\ga \rpar + O(\partial^3) =  0.
\eeq

\subsection{The Navier-Stokes Equation}
As the Navier Stokes Equation describes the conservation of momentum we take the first order velocity moment of \eref{eqn:lbeexp2} and obtain
\beq{eqn:ns1}
\sum_j v_{j\ga} \lpar \pt + v_{j\gb} \pb \rpar f_j^0 - \sum_j v_{j\ga} \lpar \pt + v_{j\gc} \pc \rpar\sum_i \lbk \lpar \Lambda^{-1}\rpar_{ji} - \frac{1}{2} \delta_{ji}\rbk \lpar \pt + v_{i\gb} \pb \rpar f_i^0 + O(\partial^3) = 0.
\eeq
%This is the same expansion as \eref{eqn:cont3} carried to second order and 
The collision term vanishes according to \eref{eqn:cons2}. We can rewrite the first of the second order terms as
\beq{eqn:ns2}
\pt \sum_j v_{j_\ga} \sum_i \lbk \lpar \Lambda^{-1}\rpar_{ji} - \frac{1}{2} \delta_{ji}\rbk \lpar \pt + v_{i\gb} \pb \rpar f_i^0 = O(\partial^3),
\eeq
which vanishes to third order due to~\eref{eqn:lbe22} much like~\eref{eqn:cont1b}. To evaluate the remaining gradient term
\beq{eqn:ns22}
\pc \sum_j v_{j\ga} v_{j\gc} \sum_i \lbk \lpar \Lambda^{-1}\rpar_{ji} - \frac{1}{2} \delta_{ki}\rbk \lpar \pt + v_{i\gb} \pb \rpar f_i^0
\eeq
we need to know the stress moments $\sum_j v_{j\ga} v_{j\gc} \lbk \lpar \Lambda^{-1} \rpar_{ji} - \frac{1}{2} \gd_{ji}\rbk$ of the collision matrix.
%can be rewritten as
%The first term~\eref{eqn:ns2} vanishes due to~\eq{eqn:lbeexp2} much like~\eref{eqn:cont1b} did.
%Because of~\eref{eqn:tj} the right hand side of \eref{eqn:ns2} is the time derivative of the right hand side of \eref{eqn:cont3} multiplied by $(\tau_{j_\ga} - 1/2)$. Since it contains a $\tau_j$ term it cannot appear in the final equations anyway. 
%As this term is of second order its time derivative is a third order term. 
%The remaining second order terms are then 
%\beq{eqn:ns3}
%\pc \sum_j  v_{j\ga} v_{j\gc}\sum_i \lbk \lpar \Lambda^{-1}\rpar_{ji} - \frac{1}{2} \delta_{ji}
%\rbk \pt f_i^0 + \pc \sum_j v_{j\ga} v_{j\gc}\sum_i \lbk \lpar \Lambda^{-1}\rpar_{ji} - \frac{1
%}{2} \delta_{ji}\rbk v_{i\gb} \pb f_i^0.
%\eeq
From the single-relaxation time derivation~\cite{wagner-2006} we know that these terms lead to the stress terms in the Navier-Stokes equation we wish to obtain. Because we want to distinguish between bulk and shear stress now we separate these into a trace and a traceless velocity moment
\beq{eqn:ns4}
\sum_j v_{j\ga} v_{j\gc} \Lambda_{ji} = \sum_j v_{j\gd} v_{j\gd} \frac{\gd_{\ga\gc}}{D} \Lambda_{ji}+ \sum_j \lpar v_{j\ga} v_{j\gc} - v_{j\gd} v_{j\gd} \frac{\gd_{\ga\gc}}{D}\rpar \Lambda_{ji}.
\eeq
The key requirement is now that the trace and the $\lpar D-1\rpar \lpar \frac{D}{2}+1\rpar$ elements of the traceless part are left eigenvectors of the collision matrix $\Lambda$. For the trace part we demand
%In order to be able to relax bulk and shear stress terms independently we require that the corresponding second order velocity moments are left eigenvectors of the collision matrix $\Lambda$. This allows us to obtain the Navier-Stokes equation. Our bulk eigenvalue equation is
%Since eigenvectors are orthogonal there appears a small problem here since $v_{k\gc} v_{k\gc}$ is not orthogonal to $1_k$ which we already have established as eigenvector in~\eref{eqn:rhoev}. We apply a Gram-Schmidt orthogonalization procedure to generate an orthogonal version of the bulk moment and obtain the bulk term in an orthogonal vector and a component proportional to the $1_k$ eigenvector. Our bulk term then is
%We thus perform a Gram-Schmidt orthogonalization procedure on the $v_{k\gc}v_{k\gc}$ moment and find
%It turns out $\sum_k v_{k\ga} v_{k\gc}$ is not orthogonal to $\sum_k 1_k$. After applying a Gram-Schmidt procedure to the trace term we require $\tau_B$ as Eigenvalue of $\lpar \Lambda^-1\rpar_{ki}$ to the orthogonalized bulk component of the second velocity moment
\beq{eqn:nsbulk}
%\sum_k \lpar v_{k\ga} v_{k\gc} - K 1_k \rpar \frac{\gd_{\ga\gc}}{D} \lpar \Lambda^{-1}\rpar_{ki} + \sum_k K 1_k \frac{\gd_{\ga\gc}}{D}\lpar\Lambda^{-1}\rpar_{ki} = \tau_B \lpar v_{i\ga} v_{i\gc} - K 1_i \rpar \frac{\gd_{\ga\gc}}{D} + \tau_\rho K 1_i \frac{\gd_{\ga\gc}}{D}, 
\sum_j v_{j\gd} v_{j\gd} \frac{\gd_{\ga\gc}}{D} \lpar \Lambda^{-1}\rpar_{ji} = \tau_B v_{i\gd} v_{i\gd} \frac{\gd_{\ga\gc}}{D}, 
\eeq
where $\tau_B$ is the bulk relaxation time and for the traceless part we require
%. Analogously the shear stress eigenvalue equation is
% Luckily the shear stress velocity term is orthogonal to all previously identified left eigenvectors and with $\tau_S$ as the shear stress relaxation time we find
\beq{eqn:nsshear}
\sum_j \lpar v_{j\ga} v_{j\gc} - v_{j\gd} v_{j\gd} \frac{\gd_{\ga\gc}}{D} \rpar \lpar \Lambda^{-1}\rpar_{ji} = \tau_S \lpar v_{i\ga} v_{i\gc} - v_{i\gd} v_{i\gd} \frac{\gd_{\ga\gc}}{D}\rpar,
\eeq
where the shear stress relaxation time $\tau_S$ is the eigenvalue. These eigenvalue equations for the second order velocity moments are the key property of the collision matrix that allows us to recover the Navier-Stokes equation. Because of the freedom to choose different eigenvalues for the trace and the traceless part we can obtain independent bulk and shear stresses. 

What follows is essentially the same derivation as in the single-relaxation time case~\cite{wagner-2006}, except that we now have two stress terms with associated relaxation times that need to be treated independently. We use the eigenvalue equations~(\ref{eqn:nsbulk}) and~(\ref{eqn:nsshear}) in~\eref{eqn:ns22} to replace $\Lambda^{-1}$ with the appropriate eigenvalues. The different velocity moments are substituted by the expressions in Eqs.~(\ref{eqn:rho} - \ref{eqn:v3}) and we replace $\tau_B - \frac{1}{2} = \omega_B$ and $\tau_S - \frac{1}{2} = \omega_S$. We get
\begin{eqnarray}
\nonumber
&&\;\;\;\pc \pt \sum_j \sum_i v_{j\ga} v_{j\gc} \lbk \lpar \Lambda^{-1}\rpar_{ji} - \frac{1}{2} \delta_{ji}\rbk f_i^0 + \pc \pb \sum_j \sum_i v_{j\ga} v_{j\gc} \lbk \lpar \Lambda^{-1}\rpar_{ji} - \frac{1}{2} \delta_{ji}\rbk v_{i\gb} f_i^0 \\ \nonumber 
&=&\;\;\;\pc \omega_B \lbk \pt \lpar \rho u_\gd u_\gd \frac{\gd_{\ga\gc}}{D} + \rho\theta \gd_{\ga\gc}\rpar + \theta\frac{D+2}{D} \gd_{\ga\gc}\pb \lpar  \rho u_\gb \rpar + \frac{\gd_{\ga\gc}}{D} \pb \lpar \rho u_\gd u_\gd u_\gb + Q_{\gd\gd\gb} \rpar \rbk \\ \nonumber
%&& - \pc \omega_B K \lbk \pt \rho + \pb \lpar \rho u_\gb \rpar \rbk \frac{\gd_{\ga\gc}}{D} + \pc \omega_\rho K \lbk \pt \rho + \pb \lpar \rho u_\gb \rpar  \rbk \frac{\gd_{\ga\gc}}{D} \\
&& +\pc \omega_S \bigg\lbrack \pt \lpar \rho u_\ga u_\gc - \rho u_\gd u_\gd \frac{\gd_{\ga\gc}}{D} \rpar + \pb \lpar \rho \theta \lpar u_\ga \gd_{\gb\gc} + u_\gb \gd_{\gc\ga} + u_\gc \gd_{\ga\gb} \rpar + \rho u_\ga u_\gb u_\gc + Q_{\ga\gb\gc} \rpar  \\
\label{eqn:ns5}
&&\;\;\;\;\;\;\;\;\; -\pb \lpar \theta \frac{ D+2 }{D} \gd_{\ga\gc} \rho u_\gb + \frac{\gd_{\ga\gc}}{D} \lpar \rho u_\gd u_\gd u_\gb + Q_{\gd\gd\gb} \rpar \rpar \bigg\rbrack.
\end{eqnarray}
%We are left with a few algebraic steps that are exactly equivalent to the single-relaxation time case detailed e.g. in~\cite{qinlipaper}. 
To treat the second order terms further we need two identities we obtain by looking at the first order terms of~\eref{eqn:ns1}. Inserting the moments~(\ref{eqn:rho}), (\ref{eqn:j}) and ignoring all second order terms we get 
\beq{eqn:nsid1}
\pt \lpar \rho u_\ga \rpar = - \pb \lpar \rho\theta \gd_{\ga\gb} + \rho u_\ga u_\gb  \rpar + O(\partial^2).
\eeq
Using the continuity equation~(\ref{eqn:contfinal}), we obtain the second identity 
\beq{eqn:nsid2}
\rho \pt u_\ga = -\rho u_\gb \pb u_\ga - \pb \rho\theta \gd_{\ga\gb} + O(\partial^2).
\eeq
These two identities and the continuity equation \eref{eqn:contfinal} now replace the time derivatives in \eref{eqn:ns5}
\begin{eqnarray}
\nonumber
&&\;\;\;\; \pc \omega_B \bigg\lbrace - \theta \delta_{\ga\gc} \pb \lpar \rho u_\gb \rpar - \frac{\gd_{\ga\gc}}{D} \lbk u_\gd \pb \lpar \rho \theta \gd_{\gb\gd} + \rho u_\gb u_\gd \rpar  + u_\gd \lpar \rho u_\gb \pb u_\gd + \pb \rho\theta \gd_{\gb\gd} \rpar \rbk  \\ \nonumber 
&&
\;\;\;\;\;\;\;\;\;\;\;\;\;\;\; +\;\frac{D+2}{D}\theta\gd_{\ga\gc} \pb \lpar  \rho u_\gb \rpar + \frac{\gd_{\ga\gc}}{D} \pb \lpar \rho u_\gd u_\gd u_\gb + Q_{\gd\gd\gb} \rpar\bigg\rbrace \\ \nonumber
&&+\; \pc \omega_S \bigg\lbrace - u_\gc \pb \lpar \rho\theta \gd_{\ga\gb} + \rho u_\ga u_\gb \rpar - u_\ga \lpar \rho u_\gb \pb u_\gc + \pb \rho \theta \gd_{\gc\gb} \rpar + \pb\lpar \rho u_\ga u_\gb u_\gc + Q_{\ga\gb\gc} \rpar \\ \nonumber 
&& \;\;\;\;\;\;\;\;\;\;\; +\; \frac{\gd_{\ga\gc}}{D}\lbk u_\gd \pb \lpar \rho\theta \gd_{\gb\gd}+ \rho u_\gb u_\gb \rpar  + u_\gd \lpar \rho u_\gb \pb u_\gd + \pb \rho\theta \gd_{\gb\gd} \rpar \rbk + \pb \rho\theta \lpar u_\ga \gd_{\gb\gc} + u_\gb \gd_{\gc\ga} + u_\gc \gd_{\ga\gb} \rpar \\ \nonumber
&& \;\;\;\;\;\;\;\;\;\;\; -\; \frac{\gd_{\ga\gc}}{D} \pb \lpar \rho u_\gd u_\gd u_\gb + Q_{\gd\gd\gb} \rpar \bigg\rbrace + O(\partial^3) \\ 
%& = &\;\;\;\; \omega_B \lbk \frac{2}{3D} \pa \rho \pc u_\gc - \frac{1}{D} \pa \pc \rho u_\gc u_\gd u_\gd \rbk \\  \nonumber
& = &\;\;\;\; \omega_B \lbk \frac{2}{D}\theta \pa \rho \pc u_\gc + \frac{1}{D} \pa \pc Q_{\gc\gd\gd} \rbk
\\  \nonumber
%&&+\; \omega_S \lbk \frac{1}{3}\pc \lpar \pc u_\ga + \pa u_\gc \rpar - \pb \pc \rho u_\ga u_\gb u_\gc - \frac{2}{3D} \pa \rho \pc u_\gc + \frac{1}{D} \pa \pc \rho u_\gc u_\gd u_\gd \rbk + O(\partial^3).
\label{eqn:ns6}
&&+\; \omega_S \lbk \theta \pc \lpar \pc u_\ga + \pa u_\gc \rpar + \pb \pc Q_{\ga\gb\gc} - \frac{2}{D} \theta \pa \rho \pc u_\gc - \frac{1}{D} \pa \pc Q_{\gc\gd\gd} \rbk + O(\partial^3).
\end{eqnarray}
If we now combine the first order terms \eref{eqn:nsid1} with the second order terms \eref{eqn:ns6} of the first order velocity moment of the LBE (\ref{eqn:ns1}) we find the Navier-Stokes equation
\begin{eqnarray}
\label{eqn:nsfinal}
%\pt \lpar \rho u_\ga \rpar + \pb \lpar \rho u_\ga u_\gb \rpar & = & - \pa \frac{\rho}{3} + \pa \omega_B \lbk \frac{2}{3D} \rho \pc u_\gc - \frac{1}{D} \pc \rho u_\gc u_\gd u_\gd \rbk \\ \nonumber
%&& + \pc \omega_S \lbk \frac{\rho}{3} \lpar \pc u_\ga + \pa u_\gc \rpar - \pb \rho u_\ga u_\gb u_\gc - \frac{2}{3D} \rho \pc u_\gc \gd_{\ga\gc} + \frac{1}{D} \pa \rho u_\gc u_\gd u_\gd \rbk \\ \nonumber
%&& + O(\partial^3).
%\pt \lpar \rho u_\ga \rpar + \pb \lpar \rho u_\ga u_\gb \rpar & = & - \pa \frac{\rho}{3} + \pa \omega_B \lbk \frac{2}{3D} \rho \pc u_\gc + \frac{1}{D} \pc Q_{\ga\gb\gc} \rbk \\ \nonumber
%&& + \pc \omega_S \lbk \frac{\rho}{3} \lpar \pc u_\ga + \pa u_\gc \rpar + \pb Q_{\ga\gb\gc} - \frac{2}{3D} \rho \pc u_\gc \gd_{\ga\gc} - \frac{1}{D} \pa Q_{\gc\gd\gd} \rbk \\ \nonumber
%&& + O(\partial^3).
\pt \lpar \rho u_\ga \rpar + \pb \lpar \rho u_\ga u_\gb \rpar & = & -\; \pa \rho\theta + \pa \omega_B \frac{2}{D}\theta \rho \pc u_\gc + \pc \omega_S \lbk \rho\theta \lpar \pc u_\ga + \pa u_\gc \rpar - \frac{2}{D}\theta \rho \pc u_\gc \gd_{\ga\gc} \rbk \\ \nonumber 
&&+\; \pa \omega_B \frac{1}{D} \pc Q_{\gc\gd\gd} + \pc \omega_S \lpar \pb Q_{\ga\gb\gc} - \pc \frac{1}{D} Q_{\gc\gd\gd}\rpar  + O(\partial^3)\\
& = & -\;\pa \rho \theta + \pa \mu \pc u_\gc + \pc \eta \lbrack \lpar \pc u_\ga + \pa u_\gc \rpar - \frac{2}{D} \pc u_\gc \delta_{\gac}\rbrack + O(\partial^2 Q) + O(\partial^3), 
%frac{1}{D} \pc Q_{\ga\gb\gc} \rbk \\ \nonumber
%&& + \pc \omega_S \lbk \frac{\rho}{3} \lpar \pc u_\ga + \pa u_\gc \rpar + \pb Q_{\ga\gb\gc} - \frac{2}{3D} \rho \pc u_\gc \gd_{\ga\gc} - \frac{1}{D} \pa Q_{\gc\gd\gd} \rbk \\ \nonumber
\end{eqnarray}
where $\mu = \frac{2}{D} \rho \theta (\tau_B-\frac{1}{2})$ is the bulk and $\eta = \rho \theta (\tau_S-\frac{1}{2})$ the shear viscosity.

In summary we recover the continuity and Navier-Stokes equations in a similar form as found from multi-relaxation time approaches with independently adjustable bulk and shear viscosities provided that three conditions are fulfilled: 
\begin{enumerate}
\item The first four velocity moments of the equilibrium distribution are given by Eqs.~(\ref{eqn:rho}-\ref{eqn:v3}).
\item The moments of the conserved quantity vectors $1_k$ and $v_{k\ga}$ are not altered in the collision step.
%\item The scalar products of the conserved moments $1_k$ and $v_{k\ga}$ are again linear combinations conserved moments. 
%The conserved moments $1_k$ and $v_{k\alpha}$ are transformed into the conserved subspace. 
\item The collision matrix has the left eigenvectors $v_{k\ga}v_{k\gb}-v_{k\gc}v_{k\gc}\frac{\gd_{\ga\gb}}{D}$ and $v_{k\gc}v_{k\gc}$. 
\end{enumerate}
Unfortunately none of the published multi-relaxation time lattice Boltzmann methods~\cite{dhumieres-1992, benzi-1992} fullfil this last requirement. This is because we have some additional freedom in combining the $\psi_i^a$ vectors with vectors from the conserved quantities as we will explain below. It is interesting to note that we have constraints up to the third order velocity moments for the equilibrium distribution, but only up to second order moments for the collision matrix. 

We should mention that the derivation presented here does not impose any requirements on the extra degrees of freedom that are typically present in a lattice-Boltzmann implementation. A D$D$Q$Q$ simulation with a $Q$ component base velocity set in $D$ dimensions only requires $ K = 1+D+D(D+1)/2 $ base vectors to reproduce isothermal hydrodynamics: $1$ for the density, $D$ for the momentum components, and $D(D+1)/2$ for the stress tensor. Our derivation makes no assumptions about the structure of the remaining $Q-K$ 'ghost' or kinetic modes or the choice of their corresponding relaxation times. Often the relaxation times for these ghost degrees of freedom are uniformly set to 1. In this case all possible choices for ghost eigenvectors of the collision matrix lead to identical collision matrices. The choice of ghost modes can influence the performance of the LB method if one wants to make use of the freedom to choose arbitrary relaxation times\cite{lallemand-2000}. The introduction of fluctuations to the LBM requires careful treatment of the ghost modes and their relaxation times \cite{adhikari-2005}, particularly in the context of boundary conditions \cite{duenweg-2007}. Furthermore Adhikari and Succi suggested a duality between conserved quantity vectors and ghost modes \cite{adhikari-2008} as guideline for constructing base velocity sets for multi-relaxation time implementation.
%\begin{verbatim}
%Issues:
%- In \enumerate: scalar products -> linear combinations -> what?
%- Q-Terms
%\end{verbatim}

\subsection{Limited Freedom of Choice of the Eigenvectors}
When we required~Eqs.~(\ref{eqn:nsbulk}) and~(\ref{eqn:nsshear}) we ignored that there is a remaining freedom of choice for the eigenvectors. To understand this, let us first remember that the relaxation  times for the conserved moments $\tau_\rho$ and $\tau_{j_\ga}$ are entirely arbitrary by construction. Because the conserved moments of the $f_i$ and $f_i^0$ are identical the collision term simply can not alter the values of the conserved quantities, independent of the value of $\tau_\rho$ and $\tau_{j_\ga}$.
This also implies that the effect of adding multiples of a conserved mode eigenvector $\psi_j^{a,c}$ to any of the eigenvectors will still result in suitable eigenvectors. Consider an alternative collision matrix $\hat{\Lambda}$ with a left eigenvector $\lpar \psi_j^n + \psi_j^c \rpar$:
\beq{eqn:ev1}
\sum_j \lbk\lpar \psi_j^n + \psi_j^c\rpar - \psi_j^c\rbk \lpar\hat{\Lambda}^{-1}\rpar_{ji} = \tau^n \lpar \psi_i^n + \psi_i^c \rpar - \tau^c \psi_i^c .
\eeq
Here $\psi_j^n$ is an eigenvector of the original matrix $\Lambda^{-1}$. The $n$ indicates that it corresponds to a non-conserved quantity and $\tau^n$ is the associated eigenvalue. In contrast $\psi_j^c$ is an eigenvector that corresponds to a conserved quantity, i.e. $\rho$ or $v_{\ga}$, with the associated eigenvalue $\tau^c$. 
%\begin{verbatim}
%Notation question: Better to use \psi_j^{a,c} as used earlier to 
%indicate that there are multiple conserved and non-conserved EVs?
%\end{verbatim}
Now, terms that depend on $\tau^c$ have to vanish because its value is entirely arbitrary. Therefore we will only retain the $\tau^n \psi^n_i$ terms in the hydrodynamic equations. The collision matrices $\Lambda$ and $\hat{\Lambda}$ will lead to identical hydrodynamic equations. To illustrate this we reinvestigate the bulk stress component in the second order terms in the Navier-Stokes derivation in ~\eref{eqn:ns5} for the alternative collision matrix $\hat{\Lambda}$. We replace $v_{j\gd}v_{j\gd}$ with $(v_{j\gd}v_{j\gd} + K 1_j)-K 1_j$ and use the aforementioned new collision matrix $\hat{\Lambda}^{-1}$ and obtain
%order velocity moment of the LBE as one would to obtain the Navier-Stokes equation and following the Navier-Stokes derivation there is no difference up to~\eref{eqn:ns22}. If we chose to replace the bulk stress eigenvector $v_{j\gd}v_{j\gd}$ by $(v_{j\gd}v_{j\gd} + K 1_j)$ the second order terms would then be
%\begin{eqnarray}
%\nonumber
%&&\pc \pt \sum_i \sum_j v_{j\ga} v_{j\gc} \lbk \lpar \hat{\Lambda}^{-1}\rpar_{ji} - \frac{1}{2} \delta_{ji}\rbk f_i^0 + \pc \pb \sum_i \sum_j v_{j\ga} v_{j\gc} \lbk \lpar \hat{\Lambda}^{-1}\rpar_{ji} - \frac{1}{2} \delta_{ji}\rbk v_{i\gb} f_i^0 \\ \nonumber 
%&=& \pc \omega_B \lbk \pt \lpar \rho u_\gd u_\gd \frac{\gd_{\ga\gc}}{D} + \rho\theta \gd_{\ga\gc}\rpar + \frac{D+2}{D}\theta \gd_{\ga\gc}\pb \lpar  \rho u_\gb \rpar \rbk \\ \nonumber
%\label{eqn:contcancel}
%&& + \pc  \omega_S \lbk \pt \lpar \rho u_\ga u_\gc - \rho u_\gd u_\gd \frac{\gd_{\ga\gc}}{D} \rpar + \pb \lpar \rho\theta \lpar u_\ga \gd_{\gb\gc} + u_\gb \gd_{\gc\ga} + u_\gc \gd_{\ga\gb} \rpar + \rho\theta \frac{ D+2 }{D} \gd_{\ga\gc} u_\gb\rpar \rbk\\
%&& - \pc \omega_B K \lbk \pt \rho + \pb \lpar \rho u_\gb \rpar \rbk \frac{\gd_{\ga\gc}}{D} +\pc \omega_\rho - \frac{1}{2}\rpar K \lbk \pt \rho + \pb \lpar \rho u_\gb \rpar  \rbk \frac{\gd_{\ga\gc}}{D} .
%\end{eqnarray}

\begin{eqnarray}
\nonumber
&&\-pc \pt \sum_i \sum_j \lbk (v_{j\gd}v_{j\gd} + K 1_j)-K 1_j \rbk \lbk \lpar \hat{\Lambda}^{-1}\rpar_{ji} - \frac{1}{2} \delta_{ji}\rbk f_i^0 \\ \nonumber 
&&\;\;\;+\;\pc \pb \sum_i \sum_j \lbk (v_{j\gd}v_{j\gd} + K 1_j)-K 1_j \rbk \lbk \lpar \hat{\Lambda}^{-1}\rpar_{ji} - \frac{1}{2} \delta_{ji}\rbk v_{i\gb} f_i^0 \\ \nonumber 
&=& \pc \omega_B \lbk \pt \lpar \rho u_\gd u_\gd \frac{\gd_{\ga\gc}}{D} + \rho\theta \gd_{\ga\gc}\rpar + \frac{D+2}{D}\theta \gd_{\ga\gc}\pb \lpar  \rho u_\gb \rpar \rbk \\%\nonumber
\label{eqn:contcancel}
%&& + \pc  \omega_S \lbk \pt \lpar \rho u_\ga u_\gc - \rho u_\gd u_\gd \frac{\gd_{\ga\gc}}{D} \rpar + \pb \lpar \rho\theta \lpar u_\ga \gd_{\gb\gc} + u_\gb \gd_{\gc\ga} + u_\gc \gd_{\ga\gb} \rpar + \rho\theta \frac{ D+2 }{D} \gd_{\ga\gc} u_\gb\rpar \rbk\\ 
&&\;\;-\; \pc \omega_B K \lbk \pt \rho + \pb \lpar \rho u_\gb \rpar \rbk \frac{\gd_{\ga\gc}}{D} +\pc \omega_{\rho} K \lbk \pt \rho + \pb \lpar \rho u_\gb \rpar  \rbk \frac{\gd_{\ga\gc}}{D} .
\end{eqnarray}
For readibility we omit the $ \rho u_\ga u_\gb u_\gc + Q_{\ga\gb\gc}$ correction terms from~\eref{eqn:v3} here as no additional third order velocity moments are generated by the $1_j$ term in the new bulk viscosity eigenvector. The $1_j$ contributions lead to additional terms consisting of derivatives of the continuity equation. Since these contributions vanish to third order \eref{eqn:contfinal} the resulting Navier-Stokes equation remains unaffected. If we decided to add a first order velocity moment to one of the non-conserved eigenvectors we would find a Navier-Stokes equation instead of the continuity equation here which again vanishes to third order. We are thus free to add any vectors corresponding to our conserved quantities to the non-conserved eigenvectors. This is the degree of freedom that allows us to impose orthogonality on the eigenvectors with respect to different inner products. 

To recover the approach of d'Humieres we now need to require all of the left eigenvectors of $\Lambda_{ji}$ be orthogonal, with respect to the inner product $\sum_j \psi_j^m \psi_j^n = \gd_{nm} N^n$ where $N^n$ is the norm of the vector $\psi^n$ which need not be normalized. The only non-orthogonal left eigenvectors here are $1_j$ and $v_{j\gc}v_{j\gc}$. We remedy this by applying a Gram-Schmidt orthogonalization procedure to find the new orthogonalized bulk stress 
\beq{eqn:dhumieresbulk}
v_{j\gc}v_{j\gc} - \frac{\sum_{j'}^N 1_{j'} v_{j'\gc} v_{j'\gc}}{N} 1_j
\eeq
and thus recover d'Humieres' basis. In contrast recovering the Benzi approach requires that the eigenvectors obey orthogonality with respect to the Hermite norm: $\sum_j \psi_j^m \psi_j^n w_j = \gd_{mn}M^n$. Again only one pair of eigenvectors is not orthogonal, $1_j$ and $v_{j\gc}v_{j\gc}$. We apply the same orthogonalization procedure, however, with the new norm and thus obtain
\beq{eqn:adhikaribulk}
v_{j\gc}v_{j\gc} - \frac{\sum_{j'}^N 1_{j'} w_j  v_{j'\gc} v_{j'\gc}}{N} 1_j
\eeq
as the orthogonal bulk stress vector.
While d'Humieres' and Benzi's approaches lead to different collision matrices it is important to note that a practical implementation of the approaches is entirely identical. This is because the eigenvectors only differ by a multiple of $1_j$, which is the density eigenvector and therefore a conserved quantity eigenvector. 

Let us assume that we have two collision matrixes $\Lambda$ and $\hat{\Lambda}$ and two corresponding sets of left eigenvectors that only differ by a conserved quantity vector $\psi^a$ and $\hat{\psi}^a =  \psi^a + \psi^{c}$. Vectors with the same index $a$ correspond to the same physical quantity and will thus correspond to the same time constant $\tau^a$. The eigenvalue equations are then
\beq{eqn:oo1}
\psi^a \Lambda = \frac{1}{\tau^a} \psi^a \;\; \text{, and} \;\; \lpar \psi^a + \psi^{c} \rpar \hat{\Lambda} = \frac{1}{\tau^a} \lpar \psi^a + \psi^{c} \rpar.
\eeq
We know that conserved quantity vectors $\psi^c$ are left eigenvectors of both $\Lambda$ and $\hat{\Lambda}$ and \eref{eqn:cons2} requires that
\beq{eqn:oo2}
\sum_i \psi_i^c \sum_j \Lambda_{ij} \lpar f_j^0 - f_j \rpar = \sum_i \psi_i^c \sum_j \hat{\Lambda}_{ij} \lpar f_j^0 - f_j \rpar = 0,
\eeq
independent of the actual choice of basis.
Now the collision operators $\Omega$ and $\hat{\Omega}$ can be defined as
\beq{eqn:oo3}
\Omega = \sum_j \Lambda_{ij} \lpar f_j^0 - f_j\rpar \;\; \text{, and} \;\; \hat{\Omega} = \sum_j \hat{\Lambda}_{ij} \lpar f_j^0 - f_j\rpar.
\eeq
Operators are defined by their action on a basis. Therefore we let $\Omega$ act on an arbitrary vector chosen from its own left eigenvector basis. Using Eqs.~(\ref{eqn:oo3}), (\ref{eqn:oo1}), and (\ref{eqn:oo2}) we get
\begin{eqnarray}
\sum_i \psi_i^a \Omega_i & = & \sum_i \psi_i^a  \sum_j \Lambda_{ij} \lpar f_j^0 - f_j\rpar \\
& = & \frac{1}{\tau^a} \sum_j \psi_j^a \lpar f_j^0 - f_j\rpar \\
& = & \frac{1}{\tau^a} \sum_j \lpar \psi_j^a + \psi_j^c \rpar \lpar f_j^0 - f_j\rpar \\
& = & \sum_i \lpar \psi_i^a + \psi_i^c \rpar \sum_j \hat{\Lambda}_{ij} \lpar f_j^0 - f_j\rpar \\
& = & \sum_i \psi_i^a \sum_j \hat{\Lambda}_{ij} \lpar f_j^0 - f_j\rpar \\
& = & \sum_i \psi_i^a \hat{\Omega}_i.
\end{eqnarray}
Thus we have proven that as long as two different bases differ only by conserved quantity left eigenvectors, the collision operators are, in fact, identical.

\section{Summary}
We presented a new general formulation for the derivation of hydrodynamics. Based on the framework of generalized or multi-relaxation time formalism we performed a direct asymptotic expansion to second order of the lattice Boltzmann equation and derived the continuity and Navier-Stokes equations for the isothermal ideal gas. Our approach is general in the sense that we do not require specific knowledge of the base velocity set and equilibrium distribution function as long as the velocity moments to third order are identical to those of the continuous case and the collision does not affect the conserved quantities.  We therefore do not require an explicit multi-relaxation time representation but instead describe all physically relevant quantities in terms of left eigenvectors of a collision matrix. These left eigenvectors can again be described in terms of velocity moments and thus we maintain a representation independent of the chosen base velocity set. The eigenvalues of the collision matrix are chosen to be inverse of the relaxation time related to the physical quantity in question. Through the relaxation times associated with bulk and shear stress terms we then get direct access to the bulk and shear viscosities.

The derivation illuminates a degree of freedom in the choice of the left eigenvectors. This is rooted in the fact that the collision does not alter conserved quantities. Therefore linear combinations of conserved quantity eigenvectors can be added to non-conserved moment left eigenvectors without changing the collision operator and by extension the hydrodynamic equations. We identify this degree of freedom as source of the validity of multi-relaxation time implementations based on different inner products such as the standard vector and the Hermite norm. In fact, we show that for the simple case of isothermal hydrodynamics the collision operators of any two realizations of multi-relaxation time Lattice Boltzmann are identical provided they conserve mass and momentum and the appropriate equilibrium distribution is chosen.

\begin{acknowledgments}
The authors would like to thank Guiseppe Gonnella and Eric Foard for enlightening discussion and helpful comments. 
This work has been supported by an ND EPSCoR seed grant. 
\end{acknowledgments}

\bibliography{lbbib2}{}
\end{document}